\documentclass[12pt]{article}
\usepackage{amssymb, amsmath, amsthm, amscd, graphicx, amsfonts, 
color, vmargin}
\newtheorem{theo}{Theorem}

\newtheorem{pro}{Proposition}

\newtheorem{conj}{Conjecture}
\newtheorem{rem}{Remark}

\title{{\textsf{On the Truncation of Systems
with\\ Non-Summable Interactions~\footnote{Work supported by Swiss National 
Foundation for Science and Conselho Nacional de Desenvolvimento 
Cient\'\i fico e Tecnol\'ogico.}}}}
\author{S.Friedli~\footnote{CBPF, Rua Dr. Xavier Sigaud 150, Urca CEP 
22290-180,
Rio de Janeiro, and 
IMPA, Estrada Dona Castorina 110, Jardim Bot\^anico CEP 22460-320,
Rio de Janeiro. email: \textsf{chach@impa.br}} ,
B.N.B. de Lima~\footnote{UFMG, Av. Ant\^onio Carlos 6627,
CEP 702 30123-970, Belo Horizonte. email: 
\textsf{bnblima@mat.ufmg.br}}}
\begin{document}
\maketitle
\begin{abstract} In this note we consider long range $q$-states
Potts models on $\mathbf{Z}^d$, $d\geq 2$.
For various families of non-summable ferromagnetic
pair potentials  $\phi(x)\geq 0$, we show that there exists, for all
inverse temperature $\beta>0$, an integer $N$ such
that the truncated model, in which all interactions between spins
at distance larger than $N$ are suppressed, has at least
$q$ distinct infinite-volume Gibbs states.
This holds, in particular, for all potentials whose 
asymptotic behaviour is of the type $\phi(x)\sim \|x\|^{-\alpha}$, 
$0\leq\alpha\leq d$.
These results are obtained using simple percolation arguments.
\end{abstract}

\textbf{Keywords:} long range Potts model, percolation, 
phase transition, truncation, non-summable interaction.


\section{Introduction}
\noindent The existence of phase transitions, in lattice systems of 
equilibrium statistical mechanics,
is a mathematically well posed problem in the
canonical framework of infinite-volume Gibbs states with \emph{summable} 
interactions \cite{Geo}. For such interactions, the 
Dobrushin Uniqueness Theorem guarantees uniqueness of the Gibbs state 
at high temperature, and 
non-uniqueness at low temperature is usually obtained as a consequence of 
high sensitivity to boundary conditions. The two-dimensional Ising model, for
example, can be prepared in distinct thermodynamic states by taking the 
thermodynamic limit along sequences of boxes with different boundary 
conditions. The combination of these high and low temperature behaviours 
leads to a well defined finite critical temperature $T_c>0$.
In general, the low
temperature phases, $0<T<T_c$, are described by local 
deviations from the ground state configurations of the 
hamiltonian, at which the measure concentrates when $T\to 0$.

\noindent For \emph{non-summable} 
interactions, infinite Gibbs states and the
thermodynamic limit are not defined. Physical quantities like the 
free energy or pressure density don't
exist, due to the fact that in the limit of large volumes, the energy 
of the system grows faster than its
size. Even the use of boundary conditions, in finite volumes, poses
problems.
The pathological behaviour of  non-summable systems was well described by 
Dyson \cite{Dy}: ``\emph{When [the potential is non-summable]
there is an infinite energy-gap between the ground states and all other
states, so that the system is completely ordered at all finite
temperatures, and there can be no question of a phase transition}''.
In other words, since
the cost for flipping any given spin is infinite, the temperature, 
even very high, can never obtain to create local deviations
from the ground states, as the ones described above in the case of
 summable potentials: at any temperature the
system is ``frozen'' in one of its ground states, and no critical 
temperature can be defined. Therefore one can infer
that infinitely large lattice systems with non-summable potentials have,
independently of the method used to describe them,
trivial thermodynamic behaviour.
Nevertheless, various ways of rescaling the thermodynamic potentials 
have been considered in the litterature, in view of obtaining 
non-trivial pseudo-densities in the thermodynamic limit. See for example
\cite{HNT} for a different scaling of the free energy of gravitational and 
electrostatic particle systems, or
\cite{CT1}, \cite{TA}, \cite{CMT}, where mean field 
versions of ferromagnetic spin models with non-summable interactions 
have been studied numerically.\\

\noindent In the present note, we study non-summable systems by using a 
simpler approach which is the
following. Let $\mu_{\phi,\Lambda}$ denote the Gibbs
distribution of the system with ferromagnetic pair interaction $\phi$ in
finite volume $\Lambda$. Since nothing can be said,
in the sense of weak convergence, about the existence of
the thermodynamic limit $$\lim_{\Lambda}\mu_{\phi,\Lambda}\,,$$
we first \emph{truncate} the potential
$\phi$ by suppressing interactions between points at distance larger
 than $N$: 
\begin{equation}\label{pottronqu}
\phi_N(x):= 
\begin{cases}
\phi(x)&\text{ if }\|x\|\leq N\,,\\
0&\text{ if }\|x\|>N\,,
\end{cases}
\end{equation}
and then study the double limiting procedure
$$\lim_{N\to\infty}\lim_{\Lambda}\mu_{\phi_N,\Lambda}\,.$$
Since $\phi_N$ has finite range, the
infinite-volume truncated model
with measure
$\mu_{\phi_N}=\lim_{\Lambda}\mu_{\phi_N,\Lambda}$ is
well defined. When $\phi$ is non-summable, the 
phenomenon which we observe is the following: for \emph{any temperature, the 
measure $\mu_{\phi_N}$ becomes sensitive to boundary
conditions once $N$ is large enough (but finite)}. That is, if $\phi$
has a finite set of ground state configurations
indexed by $s$, then the measures $\mu_{\phi_N}^s=\lim_\Lambda
\mu^s_{\phi_N,\Lambda}$ obtained by taking
the thermodynamic limit along a sequence of boxes with different boundary 
condition $s$, differ once $N$ is sufficiently
large: $\mu_{\phi_N}^s\neq\mu_{\phi_N}^{s'}$ 
when $s\neq s'$. Moreover, in the limit
$N\to\infty$, each $\mu_{\phi_N}^s$ converges weakly to $\delta_s$,
the Dirac measure concentrated on the ground
state configuration $s$. This concentration phenomenon in the 
limit $N\to\infty$ \emph{at any fixed temperature}
$T>0$ is thus similar to the one discussed above for
summable interactions in the limit $T\to 0$, and is in agreement with 
Dyson's heuristic description of non-summable interactions.
Notice that in our approach, the interaction between
 any pair of spins is restored in the limit
$N\to\infty$, and no mean field rescaling is ever used.\\

\noindent In Section \ref{SPotts} we show that the scenario presented above 
indeed occurs for the ferromagnetic
Potts model ($d\geq 2$), for various families of non-summable potentials.
Our results include, for instance, 
all potentials with slow algebraic decay (see Theorem 
\ref{T1}):
\begin{equation}\label{potnonext}
\phi(x)\sim \frac{1}{\|x\|^\alpha}\quad\text{ for some }
\quad 0\leq\alpha\leq d\,,
\end{equation}
which are the usual non-summable potentials considered in 
physics.
Nevertheless, our aim is to treat general interactions, which need not be 
asymptotically regular as in \eqref{potnonext}, but can have an
irregular structure. We
also give two results for \emph{sparse} interactions, which are of independent
interest.
In Section \ref{Smethod} we give the inequality which allows to
study this problem via independent long range percolation
and then 
reformulate and prove all our results in this setting. 
As will be seen, {the proofs are simple geometric arguments}.
In Section \ref{Sfinal} we conclude with some general remarks.\\

\noindent This work originated with the study of the problem of truncation in
independent long range percolation \cite{FLS} (see
\cite{MS}, \cite{SSV}, \cite{Be} for other cases treated in the litterature).
Therefore, the results presented
in Section \ref{Smethod} give more particular cases where this 
problem can be solved.

\section{Long Range Potts Ferromagnet}\label{SPotts}

\noindent We consider the lattice $\mathbf{Z}^d$, $d\geq 2$, with the norm 
$\|x\|=\max_{k=1,\dots,d}|x_k|$.
Interactions are defined via a
\textsf{ferromagnetic potential}, which is any function
$\phi:\mathbf{Z}^d\backslash\{0\}\to[0,+\infty)$ such that
$\sup_{x\neq 0}\phi(x)<+\infty$, with the symmetry
\begin{equation}\label{symsimple}
\phi(x)=\phi(y)\quad\text{ when }\quad\|x\|=\|y\|\,.
\end{equation}
Let $N\in\mathbf{N}$.
To each potential $\phi$ can be associated a
\textsf{truncated potential} $\phi_N$,  defined as in \eqref{pottronqu}.
In the $q$-state Potts model, $q\geq 2$ is any fixed integer, and at
each site $x\in\mathbf{Z}^d$
lives a spin $\sigma_x\in\{1,2,\dots,q\}$. When $q=2$ it thus reduces
to the Ising model. Spin configurations are elements of
$\Omega=\{1,2,\dots,q\}^{\mathbf{Z}^d}$.
Consider a finite box $\Lambda_L=[-L,+L]^d\cap\mathbf{Z}^d$, $L\geq 1$. For
$\sigma\in\Omega_{\Lambda_L}=\{1,2,\dots,q\}^{\Lambda_L}$,
the \textsf{truncated Potts Hamiltonian with boundary condition}
$\eta\in\Omega$ is given by
\begin{equation}\nonumber
H_{N,\Lambda_L}^\eta(\sigma)=-\sum_{\substack{\{x, y\}\subset\Lambda_L\\ 
x\neq y}}
\phi_N(x-y)\delta(\sigma_x,\sigma_y)
-\sum_{\substack{x\in\Lambda_L, y\in\Lambda_L^c}}
\phi_N(x-y)\delta(\sigma_x,\eta_y)\,,
\end{equation}
where $\delta(a,b)=1$ if $a=b$,  $0$ otherwise.
We will mainly be interested in considering the \textsf{pure $s$ boundary 
condition}, in which $\eta_j=s$ for all $j\in\mathbf{Z}^d$. We have, with 
some abuse of notation, 
$$H^s_{N,\Lambda_L}(s)=\min_{\sigma\in \Omega_{\Lambda_L}}
H_{N,\Lambda_L}^s(\sigma)\,.$$
Therefore, we also call the pure configurations $s$ \textsf{ground 
state configurations}.
On $\Omega_{\Lambda_L}$, the
\textsf{truncated Gibbs measure at inverse 
temperature $\beta>0$ with pure $s$ boundary condition} 
is defined by:
\begin{equation}\nonumber
\mu^{\beta,s}_{\phi_N,\Lambda_L}(\sigma)
:=\frac{1}{Z^{\beta,s}_{\phi_N,\Lambda_L}}
\exp\big(-\beta H^s_{N,\Lambda_L}(\sigma)\big)\,,
\end{equation}
where $Z^{\beta,s}_{\phi_N,\Lambda_L}$ is a normalizing factor.
Let $\mathcal{F}$ be the $\sigma$-algebra on $\Omega$ generated
by cylinder events. We
consider the infinite-volume Gibbs measures $\mu^{\beta,s}_{\phi_N}$
on $(\Omega,\mathcal{F})$,
obtained by taking limits along an increasing 
sequence of boxes~\footnote{Here we extend 
$\mu^{\beta,s}_{\phi_N,\Lambda_L}$ 
to a measure on $(\Omega,\mathcal{F})$ in the standard way:
\begin{equation}\nonumber
\mu^{\beta,s}_{\phi_N,\Lambda_L}(A):=
\sum_{\substack{\sigma\in\Omega_{\Lambda_L}}}
\mu^{\beta,s}_{\phi_N,\Lambda_L}(\sigma)
1_A(\sigma\cdot s)
\quad\forall A\in\mathcal{F}\,,
\end{equation}
where the configuration $\sigma\cdot s \in\Omega$
coincides with $\sigma$  on $\Lambda_L$ and with $s$ on $\Lambda_L^c$.} (this
limit is to be understood in the sense of subsequences):
\begin{equation}\nonumber
\mu^{\beta,s}_{\phi_N}(A):=
\lim_{L\to\infty}\mu^{\beta,s}_{\phi_N,\Lambda_L}(A)\,\quad
\forall A\in\mathcal{F}\,.
\end{equation}
A {\sf phase transition occurs} in the truncated model 
if $\mu^{\beta,s}_{\phi_N}\neq \mu^{\beta,s'}_{\phi_N}$ for $s'\neq s$.\\

\noindent Let $B_N:=\Lambda_N\backslash\{0\}$.
When $\phi$ is \textsf{summable}, i.e. when
\begin{equation}\label{sumconv}
\sum_{x\neq 0}\phi(x):=\lim_{N\to\infty}\sum_{x\in B_N}\phi(x)\,
\end{equation}
exists, the untruncated Gibbs measures (with $N=\infty$) 
$\mu^{\beta,s}_{\phi}$ are well defined, and the problem of knowing if
$\mu^{\beta,s}_{\phi}\neq \mu^{\beta,s'}_{\phi}$ 
for some $s'\neq s$ depends strongly on the temperature.
When $\phi$ is not summable these measures are not defined,
and we study $\mu^{\beta,s}_{\phi_N}$ at large $N$. 
We remind that for fixed $N$, in the limit of very low 
temperature, $\beta\to\infty$,
the typical configurations of $\mu^{\beta,s}_{\phi_N}$ 
concentrate on the ground state configuration $s$ (\cite{PS1}).
When the temperature is fixed and $N$ becomes large, we observe 
essentially the same phenomenon.
In view of the argument of Dyson cited in the Introduction, it is
reasonable to believe that at any fixed $\beta>0$, 
each of the measures $\mu^{\beta,s}_{\phi_N}$
concentrates, when $N\to\infty$, 
on a single configuration, which is the ground state $s$. This is the statement
of the following conjecture.
Let $\delta_s$ denote the Dirac mass on $(\Omega,\mathcal{F})$ 
concentrated on the ground state configuration $s$, and write
$\mu^{\beta,s}_{\phi_N}\Rightarrow \delta_s$ when $\mu^{\beta,s}_{\phi_N}$
converges weakly to $\delta_s$ in the limit $N\to\infty$.
\begin{conj}\label{Conj1}$(d\geq 2)$ If $\phi\geq 0$ satisfies 
\eqref{symsimple} and is non-summable, i.e.
\begin{equation}\label{sumdiv}
\sum_{x\neq0}\phi(x)=+\infty\,,
\end{equation}
then $\mu^{\beta,s}_{\phi_N}\Rightarrow \delta_s$ for 
all $\beta>0$ and for all $s\in\{1,2,\dots,q\}$.
\end{conj}
\noindent Observe that $\mu^{\beta,s}_{\phi_N}\Rightarrow \delta_s$
implies $\mu^{\beta,s}_{\phi_N}(\sigma_0=s)\to 1$ in the limit 
$N\to\infty$, i.e. a phase transition occurs in the truncated model for 
large enough $N$.
Since the system is ferromagnetic, the sequence 
$\big(\mu^{\beta,s}_{\phi_N}(\sigma_0=s)\big)_{N\geq 1}$ 
is non-decreasing, but the fact that it converges to $1$ is not
trivial. 
Another way to formulate the conjecture is:
$\beta_{c}(\phi_N)\to 0$ when $N\to\infty$, where 
$\beta_{c}(\phi_N)$ is the critical inverse temperature of the model with 
potential $\phi_N$, i.e.
$$\beta_c(\phi_N):=\inf\big\{\beta>0:
\mu^{\beta,s}_{\phi_N}\neq \mu^{\beta,s'}_{\phi_N}\text{ for }
s\neq s'\big\}\,.$$
\noindent 
The conjecture is difficult to prove in such generality, since we don't
assume any kind of regularity on $\phi$. For example, the potential
\begin{equation}\nonumber
\phi(x)=
\begin{cases}
\epsilon>0&\text{ if }\|x\|=k!\text{ for some }k\in\mathbf{N}\,,\\
0&\text{ otherwise}\,,
\end{cases}
\end{equation}
which will enter in the family of interactions considered in
Theorem \ref{TS1a}, satisfies the hypothesis of the conjecture. 
Usual perturbation techniques, such as Pirogov-Sinai
Theory \cite{PS1}, are of no use for studying the truncated version 
of this kind of 
potential, since the domain of validity for the temperature
shrinks to zero when the range of interaction, here $N$, grows.
\begin{rem}\label{remunique}{\rm Observe that since
the Gibbs state of any one-dimensional model with finite range
interactions is always unique (see for example Theorem (8.39) in \cite{Geo}),
Conjecture \ref{Conj1} is \emph{false} in dimension $1$.
This shows that a symmetry of the kind \eqref{symsimple} is necessary.
}\end{rem}
\noindent For $d\geq 2$, our first result shows that
the conjecture is valid under some
assumption on the speed of divergence of the series \eqref{sumdiv}.
\begin{theo}$(d\geq 2)$\label{T1}
If $\phi\geq 0$ satisfies \eqref{symsimple} and
\eqref{sumdiv} diverges faster than logarithmically, i.e.
\begin{equation}\label{cond0}
\limsup_{N\to\infty}\frac{1}{\log |B_N|}\sum_{x\in B_N}\phi(x)
=+\infty\,,
\end{equation}
then $\mu^{\beta,s}_{\phi_N}\Rightarrow \delta_s$ for 
all $\beta>0$ and for all $s\in\{1,2,\dots,q\}$.
\end{theo}
\noindent 

\begin{rem}\label{R1}{\rm 
As can be seen easily,
condition \eqref{cond0} is satisfied by all potentials which have slow 
algebraic decay, as in \eqref{potnonext}:
\begin{equation}\nonumber
\liminf_{\|x\|\to\infty}\|x\|^\alpha \phi(x)>0\quad\text{ for some }
0\leq\alpha<d\,.
\end{equation}
We shall see later, in Remark \ref{R9},
that for such potentials the range $0\leq \alpha<d$
can be extended to $0\leq \alpha\leq d$, using the multiscale analysis of
\cite{Be}.
}\end{rem}
\paragraph{Sparse Interactions.}
We also give two
results for potentials which don't have the symmetry \eqref{symsimple}.
Namely, we
consider interactions only along directions parallel to the coordinate axis
$e_i$, $i=1,\dots,d$, where $e_1=(1,0,\dots,0)$, $e_2=(0,1,0,\dots,0)$, ..., 
$e_d=(0,0,\dots,1)$. 
That is, we are given a sequence 
$(\phi_n)_{n\geq 1}$, $\phi_n\geq 0$, and~\footnote{The reader should pay
attention to the following: we use $n$ to index elements of the sequence
$(\phi_n)_{n\geq 1}$, whereas $N$ is used as the parameter of truncation for
$\phi_N$.}
\begin{equation}\label{formdup}
\phi(x)=
\begin{cases}
\phi_{\|x\|}&\text{ if $x$ is parallel to some }\, e_i\,,i=1,2,\dots,d\,,\\
0&\text{ otherwise}.
\end{cases}
\end{equation}
In $d=2$, for example, there are only vertical and horizontal couplings.
For potentials of the form \eqref{formdup}, 
assumption \eqref{sumdiv} of Conjecture 
\ref{Conj1} becomes:
\begin{equation}\label{sumdivS}
\sum_{n\geq 1}\phi_n=+\infty\,.
\end{equation}
\noindent The first result is for sequences $(\phi_n)_{n\geq 1}$
which don't converge to zero:
\begin{theo}\label{TS1a}
$(d\geq 2)$. If
\begin{equation}\label{otimo}
\limsup_{n\to\infty}\phi_n>0\,,
\end{equation}
then $\mu^{\beta,s}_{\phi_N}\Rightarrow \delta_s$ for
all $\beta>0$ and for all $s\in\{1,2,\dots,q\}$.
\end{theo}
\noindent Notice that 
\eqref{otimo} implies \eqref{sumdivS}, but with no information on
the speed of divergence. Our second result is where we prove
our conjecture under the general condition \eqref{sumdivS}, 
with no assumption on the speed of divergence, but
only in dimensions three or more:
\begin{theo}\label{TS2a} 
$(d\geq 3)$. If
\begin{equation}
\sum_{n\geq 1}\phi_n=+\infty\,,
\end{equation}
then $\mu^{\beta,s}_{\phi_N}\Rightarrow \delta_s$ for 
all $\beta>0$ and for all $s\in\{1,2,\dots,q\}$.
\end{theo}

\section{Independent Long Range Percolation}\label{Smethod}
To show that the truncated $q$-states 
Potts model exhibits a phase transition, 
it is sufficient to show that
\begin{equation}\label{bornundemie}
\mu_{\phi_N}^{\beta,s}(\sigma_0=s)>\frac{1}{q}\,.
\end{equation}
Our purpose is to obtain the following stronger limiting behaviour:
\begin{equation}\label{bornundemi}
\lim_{N\to\infty}\mu_{\phi_N}^{\beta,s}(\sigma_0=s)=1\,,
\end{equation}
which is equivalent to $\mu_{\phi_N}^{\beta,s}\Rightarrow \delta_s$, as can be
verified easily.
To obtain \eqref{bornundemi} we shall reformulate our problem
in the framework of long range {independent} percolation.\\

\noindent Consider the graph $(\mathbf{Z}^d,{\cal E}^d)$, $d\geq 1$, 
where  ${\cal E}^d$ is the set of all unoriented edges $e=\{x,y\}
\subset\mathbf{Z}^d\times\mathbf{Z}^d$, $x\neq y$.
Edge configurations are elements $\omega\in\{0,1\}^{\mathcal{E}^d}$.
For a given function
$p:\mathbf{Z}^d\backslash\{0\}\to [0,1]$ with
\begin{equation}\label{blob}
p(x)=p(y)\quad\text{ when }\quad \|x\|=\|y\|\,,
\end{equation}
called \textsf{edge probability}, we
consider the \textsf{long range percolation process} 
in which each edge $e=\{x,y\}$ is 
\textsf{open} ($\omega(e)=1$) with probability $p(x-y)$, and \textsf{closed} 
($\omega(e)=0$) with probability $1-p(x-y)$, independently of other edges.
This process is described by the product measure
on the $\sigma$-field on $\{0,1\}^{\mathcal{E}^d}$ generated by cylinders,
given by
\begin{equation}\label{defprodmeas}
P=\prod_{\substack{e\in{\cal E}^d}}\mu_{e}\,,
\end{equation}
where $\mu_{e}(\omega(e)=1)=p(x-y)$ is a Bernoulli measure on
$\{0,1\}$, independent of the state of other edges.
Observe that $P$ is well-defined even when $p$ is non-summable.
Define the \textsf{truncated edge probability}
\begin{equation}\nonumber
p_N(x):=
\begin{cases}
p(x)&\text{if}\quad \|x\|\leq N\,,\\
0&\text{if}\quad \|x\|> N\,,
\end{cases}
\end{equation}
and denote by $P_N$ the truncated product measure defined as in 
\eqref{defprodmeas} with $p_N$ instead of $p$.
We shall be interested in the \textsf{percolation probability}
$P_N(0\leftrightarrow\infty)$, which 
is the probability of the event in which there exists, 
in the truncated model, a path of open edges connecting 
the origin to infinity. When
$P_N(0\leftrightarrow\infty)>0$, we say the truncated system 
\textsf{percolates}.\\

\noindent As can be seen in the following proposition, the percolation
probability is a relevant quantity for showing all
our results for the dependent Potts model, once the edge probability
is well chosen in function of the potential $\phi$. 
\begin{pro}\label{lemfond}
Define $p(x)$ by 
\begin{equation}\label{equpJ}
p(x):=\frac{1-e^{-2\beta \phi(x)}}{1+(q-1)e^{-2\beta \phi(x)}}\,.
\end{equation}
Then the magnetisation of the truncated long range $q$-states
Potts model with potential $\phi_N(x)$ at temperature $\beta$
and the probability of percolation of the origin
in the independent truncated 
long range percolation process with probabilities $p_N(x)$ 
are related by the following inequality:
\begin{equation}\label{inequfond}
\mu_{\phi_N}^{\beta,s}(\sigma_0=s)\geq\frac{1}{q}+\frac{q-1}{q}
P_N(0\leftrightarrow\infty)\,.
\end{equation}
\end{pro}
\noindent Using \eqref{bornundemie}, \eqref{inequfond}
shows that percolation in the truncated independent model
implies phase transition in the truncated Potts model. This holds once
the potential $\phi$ and the probability $p$ are related by \eqref{equpJ},
which is well suited for our purposes since \eqref{equpJ} implies that
$\phi$ and $p$ are bound to have {the same asymptotic behaviour}: as can
be seen, there exist two positive functions $C_\pm=C_\pm(\beta,q)$ 
such that
\begin{equation}\nonumber
C_-\phi(x)\leq p(x)\leq C_+\phi(x)\quad\forall x\neq 0\,.
\end{equation}
In particular, non-summability of $\phi$ implies non-summability of 
$p$. Although it does not appear exactly in this form in
the litterature, \eqref{inequfond} is standard, and can be obtained 
via the random cluster representation of the measure $\mu_{\phi_N}^{\beta,s}$, 
and its domination properties with respect to Bernoulli product measures. 
We refer to \cite{ACCN} or \cite{GHM} for details.\\

\begin{figure}[htb]
\begin{center}
\input{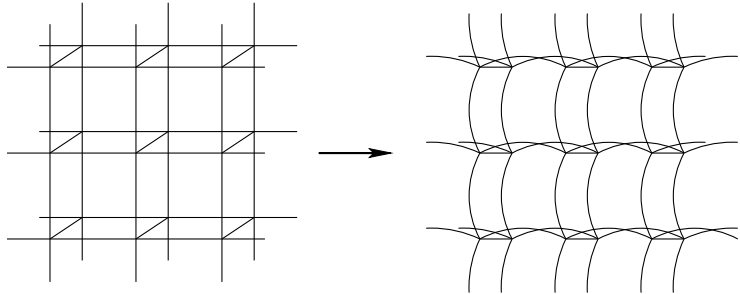}
\end{center}
\caption{\footnotesize{
The embedding of the
slab $\{1,2\}\times \mathbf{Z}^2\subset \mathbf{Z}^3$ in
$(\mathbf{Z}^2,\mathcal{E}^2)$.}}
\label{simpleslab}
\end{figure}

\begin{figure}[htb]
\begin{center}
\input{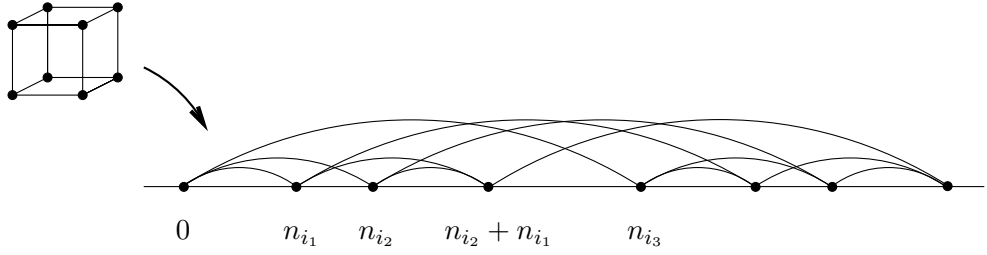}
\end{center}
\caption{\footnotesize{The embedding of the cube $\{1,2\}^3$ on
the positive axis of $(\mathbf{Z}^2,\mathcal{E}^2)$, using long range 
edges. We have
chosen $n_{i_1}:=n_1$, $n_{i_2}:=\min\{n_k:n_k>n_{i_1}\}$, and
$n_{i_3}:=\min\{n_k:n_k>n_{i_1}+n_{i_2}\}$, in order to avoid overlapping. 
Then, an infinite number
of copies of this cube must be glued together, to form the slab
$\{1,2\}^{3}\times\mathbf{Z}^{2}$ embeded in $(\mathbf{Z}^2,\mathcal{E}^2)$.}}
\label{cube}
\end{figure}

\noindent By Proposition \ref{lemfond}, long range
independent percolation can be used to show all the 
results stated previously for the Potts model.
We therefore reformulate and prove 
the equivalent of the results of Section \ref{SPotts}
in the context of independent percolation, in which our conjecture is that
when \eqref{blob} holds and 
\begin{equation}\label{lepaussi}
\sum_{x\neq 0}p(x)=+\infty\,,
\end{equation}
then $\lim_{N\to\infty}P_N(0\leftrightarrow \infty)=1$.
A simple use of the Borel-Cantelli Lemma 
shows that \eqref{lepaussi} implies $P(0\leftrightarrow \infty)=1$.
Therefore the conjecture is that for non-summable edge probabilities, 
$\lim_{N\to\infty}P_N(0\leftrightarrow \infty)=
P(0\leftrightarrow \infty)$.
Our first result for percolation is
\begin{theo}\label{T1b}
$(d\geq 2)$ There exists $c=c(d)>0$ such that if \eqref{blob} holds and 
if
\begin{equation}\label{cond0cd}
\limsup_{N\to\infty}\frac{1}{\log |B_N|}\sum_{x\in B_N}p(x)\geq c\,,
\end{equation}
then $\lim_{N}P_N(0\leftrightarrow \infty)=1$.
\end{theo}
\begin{rem}\label{R9}
{\rm As can be seen easily,
\eqref{cond0cd} is satisfied when
\begin{equation}\label{lastar}
\lambda:=\liminf_{\|x\|\to\infty}\|x\|^\alpha p(x)>0\,,\quad\text{ for some }
\quad 0\leq \alpha<d\,.
\end{equation}
In fact Theorem \ref{T1b} can be obtained, under condition \eqref{lastar},
using the multiscale analysis of Berger \cite{Be}, for all 
$0\leq \alpha \leq d$. Our Theorem allows $p$ to have gaps, but does not
cover the case $\alpha=d$, unless $\lambda$ is assumed sufficiently 
large.}
\end{rem}
\begin{proof}[Proof of Theorem \ref{T1b}:]
The proof is a simple (almost trivial) blocking argument.
Fix $N$ large. For each $s\in S:=\{(s_1,\dots,s_d):s_i=\pm 1\}$,
define the quadrant $Q_s^N(0):=\{y\in\mathbf{Z}^d:
0<s_iy_i\leq N,\,i=1,2,\dots,d\}$.
For any $x\in\mathbf{Z}^d$, let $Q_s^N(x):=x+Q_s^N(0)$.
A site $x$ is \textsf{good} if there exists, for all $s\in S$, a site 
$y\in Q_s^N(x)$ such that the edge $\{x,y\}$ is open (see Figure \ref{good}). 
We have 
\begin{align}
P_N(x\text{ is good})
&=\prod_{s\in S}\Big[1-\prod_{y\in Q_s^N(x)}
(1-p(x-y))\Big]\nonumber\\
&=\Big[1-\prod_{y\in Q_s^N(0)}
(1-p(y))\Big]^{|S|}\quad\forall s\in S\nonumber\\
&\geq  \Big[1-\exp\Big(-\sum_{y\in Q_s^N(0)}p(y)\Big)
\Big]^{|S|}\quad\forall s\in S\nonumber\\
&\geq  \Big[1-\exp\Big(-c_1\sum_{y\in B_N}p(y)\Big)
\Big]^{|S|}\,,\label{labelle}
\end{align}
where we used the inequality $\log(1-t)\leq -t$ valid for all $t<1$, and 
$c_1>0$ is a constant that depends only on the dimension. 
Next, consider a partition of $\mathbf{Z}^d$ into disjoint 
blocks of linear size $3N$, obtained by translates of the block 
$C^N(0):=[0,3N)^d\cap\mathbf{Z}^d$. That is, each block of this partition 
is of the form $C^N(z)=3zN+C^N(0)$, for some renormalized vertex 
$z\in\mathbf{Z}^d$. We say a block $C^N(z)$ is \textsf{good} if each 
$x\in C^N(z)$ is good.
\begin{figure}[h]
\begin{center}
\input{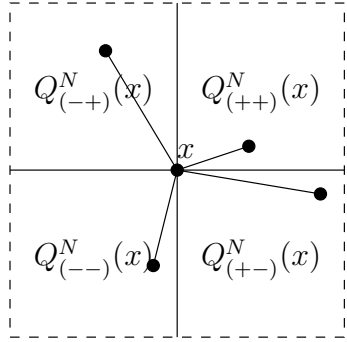}
\end{center}
\caption{\footnotesize{Illustration of a good site $x$ in the two-dimensional
case.}}
\label{good}
\end{figure}
Now for two points $x\neq x'$, the events 
$\{x\text{ is good}\}$ and $\{x'\text{ is good}\}$ are not necessarily
independent, but since they are increasing, FKG inequality gives
\begin{align}\nonumber
P_N(C^N(z)\text{ is good})&\geq\prod_{x\in C^N(z)}P_N(x\text{ is good})\,.
\end{align}
For small enough $\epsilon>0$ we have $1-\epsilon\geq e^{-2\epsilon}$. Using 
$|C^N(z)|\leq c_2|B_N|$ and 
\eqref{labelle} we thus get, when $N$ is large enough,
\begin{align}
P_N(C^N(z)\text{ is good})&\geq
\exp\Big[-2c_2|S||B_N|\exp\Big(-c_1\sum_{y\in B_N}p(y)\Big)
\Big]\nonumber\\
&=\exp\Big[-2c_2|S|\exp\Big(\log|B_N|-c_1\sum_{y\in B_N}p(y)\Big)
\Big]\,.\nonumber
\end{align} 
If \eqref{cond0cd} holds with a well chosen constant $c>0$
we get
\begin{equation}
\limsup_{N\to\infty}P_{N}(C^N(z)\text{ is good})=1\,,\nonumber
\end{equation}
i.e. there exists a sequence
$\delta_k\searrow 0$ and a diverging sequence 
$N_1,N_2,\dots$ such that 
\begin{equation}\label{bornfifi}
P_{N_k}(C^{N_k}(z)\text{ is good})\geq 1-\delta_k\,\,\forall k\,.
\end{equation}
Notice that for each $k$
the process 
$(X^k_z)_{z\in\mathbf{Z}^d}$ defined by
\begin{equation}
X^k_z:=
\begin{cases}
1&\text{ if $C^{N_k}(z)$ 
is good}\,,\\
0&\text{ otherwise}\,,
\end{cases}
\end{equation}
is \textsf{1-dependent}, i.e. $X^k_z$ and 
$X^k_{z'}$ are independent when $\|z-z'\|>1$.
By a Theorem of Liggett, 
Stacey and Schonmann \cite{LSS}, $(X^k_z)_{z\in\mathbf{Z}^d}$ 
stochastically dominates an independent Bernoulli process 
$(Z^k_z)_{z\in\mathbf{Z}^d}$ of parameter $\rho_k>0$, and
\eqref{bornfifi} implies $\lim_{k\to\infty}\rho_k=1$. 
Take $k$ large enough such 
that $\rho_k>p_c(\mathbf{Z}^d,\text{site})$, where 
$p_c(\mathbf{Z}^d,\text{site})$ is the critical threshold of 
Bernoulli nearest-neighbour site percolation. For such $k$, the process
$(Z^k_z)_{z\in\mathbf{Z}^d}$ is supercritical and by domination 
there exists an infinite
cluster of good boxes. It is easy to see that any infinite 
connected component of good boxes yields an infinite connected component of 
sites of the original lattice. By taking $N$ large one can thus make
$P_N(0\leftrightarrow \infty)$ arbitrarily close to $1$.
\end{proof}
\begin{rem}\label{R10}{\rm
Criterium \eqref{cond0cd} concerns
the behaviour of the sum
$\sum_{x\in B_N}p(x)$ for large $N$, \emph{and not the details of the function
$p(\cdot)$}. This has an interesting consequence, as the following discussion
shows. Assume $p$ is such that \eqref{cond0cd} holds. 
Then Theorem \ref{T1b} guarantees the existence of some $N$ such that 
the system 
whose edges $e=\{x,y\}$ are all of size at most $\|x-y\|\leq N$, 
with edge probabilities $p(\cdot)$, satisfies 
$P_N(0\leftrightarrow\infty)>0$. As seen in the proof,
the integer $N$ is fixed once the sum of the edge probabilities passes a
given value $K_N$, i.e.
$$\sum_{x\in B_N}p(x)\geq K_N\,.$$
Now, observe that the function $p$
can be modified inside $B_N$, but as long as the \emph{sum} is preserved,
the percolation probability remains positive. For example, if
$\pi:B_N\to B_N$ is any permutation preserving the symmetry 
$\|\pi(x)\|=\|\pi(y)\|$ for $\|x\|=\|y\|$, then 
$$\sum_{x\in B_N}p(\pi(x))=\sum_{x\in B_N}p(x)\geq K_N\,,$$
and so the truncated system with edge probability 
$p(\pi(\cdot))$ also percolates. This comment suggests that the sum 
$\sum_{x\in B_N}p(x)$, rather than the individual edge 
probabilities, is the relevant parameter in the study
of percolation in the truncated model. This property can also be 
easily verified for long range percolation on trees.}
\end{rem}
\paragraph{Sparse Connections.}
We now give two results concerning systems where connections are not 
isotropic, and we will consider the case where a connection can be opened 
between two sites $x,y$ only if these lie on a same coordinate axis.
That is, $p(x)\neq 0$ only if $x$ is parallel to one of 
the coordonate axis $e_i$, $i=1,\dots,d$. 
In this case the probabilities $p(x)$ are determined by 
a sequence $(p_n)_{n\geq 1}$, $p_n\in[0,1]$, and 
\begin{equation}\label{blub}
p(x)=
\begin{cases}
p_{\|x\|}&\text{ if $x$ is parallel to some }\, e_i\,, i=1,2,\dots,d\,,\\
0&\text{ otherwise}.
\end{cases}
\end{equation}
We expect that 
\begin{equation}\label{elledivergeoui}
\sum_{n\geq 1}p_n=+\infty\,
\end{equation}
implies $\lim_NP_N(0\leftrightarrow\infty)=1$.
\noindent A particular case of \eqref{elledivergeoui} 
in which nothing is assumed 
about the speed of divergence of the series is the following.
\begin{theo}\label{TS1b}
$(d\geq 2)$
If 
\begin{equation}\label{condtotal}
\limsup_{n\to\infty}p_n>0
\end{equation}
then $\lim_NP_N(0\leftrightarrow\infty)=1$.
\end{theo}
\noindent Since this result 
has already appeared in \cite{FLS} we shall only remind the 
strategy of the proof, which is very different, in spirit, from that
of Theorem \ref{T1b}. 
We consider the two dimensional case $d=2$. 
Nevertheless, the core of the proof is to use properties of 
nearest-neighbour percolation in high dimensions $d_*$.
Denote
by $p_c(\mathbf{Z}^{d_*})$ the percolation threshold of nearest-neighbour
Bernoulli edge percolation on $\mathbf{Z}^{d_*}$. It was shown by Kesten
\cite{Ke1} that 
\begin{equation}\label{eqKesten}
p_c(\mathbf{Z}^{d_*})\to 0\,\text{ when }\,d_*\to\infty\,.
\end{equation}
Then, let $\{1,2,\dots,L\}^{d_*-2}\times\mathbf{Z}^{2}$ denote the slab
of thickness $L$ in $\mathbf{Z}^{d_*}$.
It was shown by Grimmett and Marstrand \cite{GM} that 
the slab percolation threshold satisfies
\begin{equation}\label{eqGM}
p_c(\{1,2,\dots,L\}^{d_*-2}\times\mathbf{Z}^{2})\to
p_c(\mathbf{Z}^{d_*})\,\text{ when }\,L\to\infty\,.
\end{equation}
Now call $2\epsilon$ the $\limsup$ in \eqref{condtotal}, and 
consider some diverging
sequence $n_1,n_2,\dots$ for which $p_{n_k}\geq \epsilon$. 
By \eqref{eqKesten} and \eqref{eqGM} 
there exists a dimension $d_*$ and an integer $L$ (both 
depending on $\epsilon$) such that for all $k$, 
\begin{equation}\label{rhai}
p_{n_k}\geq \epsilon>p_c(\{1,2,\dots,L\}^{d_*-2}\times\mathbf{Z}^{2})\,.
\end{equation}
It is then clear how to pursue: we
\emph{embed} the
slab $\{1,2,\dots,L\}^{d_*-2}\times\mathbf{Z}^{2}$ in
$(\mathbf{Z}^2,\mathcal{E}^2)$, using edges of
sizes taken in the set $\{n_1,n_2,\dots\}$. Here, $N$ must be taken large
enough. The point is that we only 
need a \emph{finite}
number of sizes, and that by \eqref{rhai} each edge of the 
embedded graph has probability at least 
$\epsilon>p_c(\{1,2,\dots,L\}^{d_*-2}\times\mathbf{Z}^{2})$ of being open. 
This guarantees that the independent truncated 
process on this graph is supercritical, hence contains with probability 
one an infinite cluster: $P_N(0\leftrightarrow\infty)>0$.
The simplest case for which the embedding can be easily understood is
when $d_*=3$ and $L=2$, which we illustrated on Figure \ref{simpleslab}.
For higher dimension, for example in the case where $d_*=5$,
we have represented on Figure \ref{cube}
an embedding of the cube $\{1,2\}^3$ in $(\mathbf{Z}^2,\mathcal{E}^2)$, 
which shows what must be done in the general case. 
The formal
embedding of the slab $\{1,2,\dots,L\}^{d_*-2}\times\mathbf{Z}^{2}$
can be found in \cite{FLS}. We leave it as an exercise to the reader to show
that \eqref{eqKesten} can be used again to show that
$\lim_NP_N(0\leftrightarrow \infty)=1$.\\

\begin{rem}{\rm In order to deduce Theorem \ref{TS1a} from
Theorem \ref{TS1b}, one must estimate
\begin{align}\nonumber
\limsup_{n\to\infty}p_n
=\limsup_{n\to\infty}
\frac{1-e^{-2\beta\phi_n}}{1+(q-1)e^{-2\beta\phi_n}}
\geq C_-(\beta,q)\limsup_{n\to\infty} \phi_n\,.
\end{align}
Observe that $C_-(\beta,q)$ goes to zero when $\beta\searrow 0$. Therefore, 
the dimension $d_*$ and the thickness $L$ 
of the slab used in the proof above diverges
for large temperatures.}
\end{rem}
\begin{rem}{\rm 
Theorem \ref{TS1a} follows from Theorem \ref{TS1b} and
Proposition \ref{lemfond}. 
Recently, Bodineau \cite{Bo} has proved the analog of the result of Grimmett and
Marstrand \cite{GM} (which we used in 
\eqref{eqGM}) for the percolation threshold
of the random cluster measure associated to the Ising model ($q=2$). Therefore,
using the main result of \cite{Bo} and the same embedding as above, one can
obtain a more direct proof of Theorem \ref{TS1a} for the case $q=2$, 
without using Proposition \ref{lemfond}.}
\end{rem}
\begin{theo}\label{T3abc}$(d\geq 3)$
If 
\begin{equation}\label{ahouiellediverge}
\sum_{n\geq 1}p_n=+\infty
\end{equation}
then $\lim_NP_N(0\leftrightarrow\infty)=1$.
\end{theo}
\begin{proof}
The proof relies on the fact that sequences $p_n$ that satisfy
\eqref{ahouiellediverge} already imply percolation in one dimension.
So for a while we consider long range percolation on 
$(\mathbf{Z},\mathcal{E}^1)$, and denote
by $P^{1}$ the product measure on $\{0,1\}^{\mathcal{E}^1}$
associated to the sequence $(p_n)_{n\geq 1}$.
By Borel-Cantelli we have $\theta:=P^1(0\leftrightarrow\infty)=1$.
For simplicity we first assume that
there exists, for all $m\in\mathbf{Z}$,
a sequence $n_1=0,n_2,\dots,n_k=m$ such that
$p_{|n_{i+1}-n_i|}>0$.
This implies, by \cite{AKN}, that the infinite cluster is unique. We can 
thus write
\begin{align}
P^1(0\leftrightarrow 1)&\geq 
P^1(0\leftrightarrow\infty,1\leftrightarrow\infty)\label{yoo}\\
&\geq  P^1(0\leftrightarrow
\infty)P^1(1\leftrightarrow\infty)=\theta^2\,,\nonumber
\end{align}
where we have used, in this order, uniqueness of the infinite cluster, FKG
inequality, and translation invariance. For $a<b$ let
$T_L(a,b):=[a-L,b+L]$. Then $\{0\leftrightarrow 1\,\mathrm{in}\,
T_L(0,1)\}\nearrow
\{0\leftrightarrow 1\}$ when $L\to\infty$. Therefore for all $\delta>0$ there
exists $L_\delta$ such that for $L\geq L_\delta$,
\begin{equation}\label{PT21}
P^1(0\leftrightarrow 1 \,\mathrm{in}\, T_L(0,1))\geq \theta^2-\delta\,.
\end{equation}
Since 
\begin{equation}\label{PT22}
\theta^2=1>p_c({\mathcal{H}})\,,
\end{equation}
where $p_c({\mathcal{H}})$ is
the percolation threshold of the honeycomb lattice $\mathcal{H}$, we can
take $\delta$ small enough such that
$\theta^2-\delta>p_c({\mathcal{H}})$, and fix $L\geq L_\delta$.

\noindent Back to $d=3$, 
consider the embedding of $\mathcal{H}$ in
$\mathbf{Z}^3$, denoted $\mathcal{H}'$, depicted on Figure \ref{honeycomb}. 
\begin{figure}[htb]
\begin{center}
\input{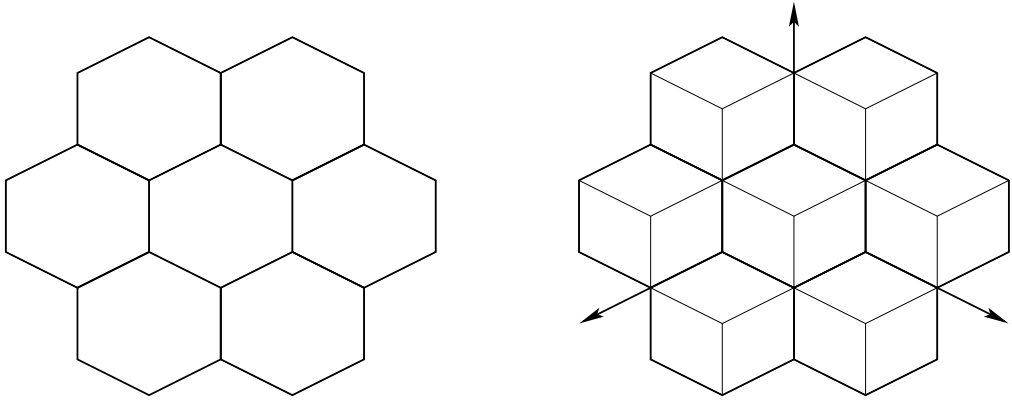}
\end{center}
\caption{\footnotesize{The honeycomb lattice 
$\mathcal{H}$ and its embedding in $\mathbf{Z}^3$, denoted $\mathcal{H}'$.}}
\label{honeycomb}
\end{figure}
To each edge $e=\{x,y\}\in\mathcal{H}'$
corresponds a one dimensional subgraph $(\mathbf{Z}(e),\mathcal{E}^1(e))
\subset (\mathbf{Z}^3,\mathcal{E}^3)$ which consists of all the
points (and long range edges)
contained in the line supported by $e$; $(\mathbf{Z}(e),\mathcal{E}^1(e))$ 
is nothing
but a copy of $(\mathbf{Z},\mathcal{E}^1)$, embedded in 
$(\mathbf{Z}^3,\mathcal{E}^3)$, containing $e$.
All the previous considerations on $(\mathbf{Z},\mathcal{E}^1)$ 
(for example the intervals $T_L(x,y)$) can be 
adapted in each subgraph $(\mathbf{Z}(e),\mathcal{E}^1(e))$.
An important property of the embedding we have chosen
is that \emph{the graphs $(\mathbf{Z}(e),\mathcal{E}^1(e))$, 
$(\mathbf{Z}(e'),\mathcal{E}^1(e'))$
associated to two different edges $e,e'\in \mathcal{H}'$ 
have disjoint sets of edges}.
We now define an edge $e=\{x,y\}\in\mathcal{H}'$ to be
\textsf{good} if and only if
there exists, in $(\mathbf{Z}(e),\mathcal{E}^1(e))$, 
a path connecting $x$ to $y$
in $T_L(x,y)$. Clearly, edges are good independently and for $N= 2L+2$,
\begin{equation}\nonumber
P_N(e\mathrm{\,\, is\,\, good})=P^1_N(x\leftrightarrow y \,
\mathrm{in}\, T_L(x,y))
\geq \theta^2-\delta>p_c(\mathcal{H})\,.
\end{equation}
Therefore, there exist infinite paths of good edges, 
yielding the existence of an infinite cluster on
$(\mathbf{Z}^3,\mathcal{E}^3)$ with edges of sizes smaller than $N$.
Clearly, $\lim_NP_N(0\leftrightarrow\infty)=1$, which finishes the proof.
When the assumption made at the beginning is not satisfied, it suffices to
replace $P^1(0\leftrightarrow 1)$, in \eqref{yoo}, by 
$P^1(0\leftrightarrow K)$ for a well chosen $K$. The rest of the proof can be
adapted in a straightforward way.
\end{proof}
\noindent Observe that the only place where we used the 
divergence of the series 
\eqref{ahouiellediverge} was to obtain $P^1(0\leftrightarrow\infty)=1$.
A variant of Theorem \ref{T3abc} can therefore be reformulated under a
more abstract condition on the sequence $(p_n)_{n\geq 1}$, which can hold also
when the series $\sum_np_n$ converges:
\begin{theo} $(d\geq 3)$. Assume the sequence $(p_n)_{n\geq 1}$ is such that
$\theta:=P^1(0\leftrightarrow \infty)$ satisfies $\theta^2>p_c(\mathcal{H})$ 
and that the one-dimensional infinite cluster is unique. Then 
$P_N(0\leftrightarrow \infty)>0$ when $N$ is large enough.
\end{theo}
\section{Final Remarks}\label{Sfinal}

We have considered the problem of truncation
in the long range Potts model with non-summable
ferromagnetic interactions, via simple percolation techniques. 
We have shown that for various families of potentials, 
a phase transition occurs in the truncated 
model as soon as the parameter of truncation $N$ is taken sufficiently 
large. Notice that by Proposition \ref{lemfond} all 
the existing results on truncation in long range percolation
(\cite{MS, SSV, Be}) have their counterpart in the long range Potts
model.
We hope that
\emph{embeddings}, as those we used in the proofs of 
Theorems \ref{TS1b} and \ref{T3abc}, might be used for possible
generalisations since they don't require any particular 
regularity of the potential/edge probability at
infinity, and give some insight into new mechanisms of phase 
transitions in systems with long range interactions.\\

\noindent Before ending, we make two remarks concerning 
the problem of truncation.

\paragraph{A Mean Field Limit.} 
As our results show, \emph{infinite} systems with non-summable
interactions have trivial dependence on the temperature. 
In the physics litterature, some methods have been used in
view of understanding the properties of large but \emph{finite} 
systems with non-summable interactions. These methods rely
essentially on the study of the mean field version of the original
model. Namely, since the energy of the finite
system grows faster than its size in the limit of large volumes
$\Lambda$, the model is modified (\cite{CMT, CT1, TA}) by dividing
the total hamiltonian by a well chosen power of the
volume $|\Lambda|$. In the case of Ising spins
$\sigma_x=\pm 1$ with ferromagnetic 
interactions $\phi(x)=\|x\|^{-\alpha}$, $0\leq \alpha\leq
d$, this means
considering the following formal identity:
\begin{equation}\label{lechampmou}
\beta\sum_{\substack{x\neq y}}\phi(x-y)\sigma_x\sigma_y=
\beta |\Lambda|^\delta
\sum_{\substack{x\neq y}}
\frac{\phi(x-y)}{|\Lambda|^\delta}\sigma_x\sigma_y\,. 
\end{equation}
The scaling parameter
$\delta$ must be chosen in function of $\alpha$ in order to obtain
a well defined thermodynamic limit for the potential 
$|\Lambda|^{-\delta}\phi(x-y)$, leading to a mean field inverse
critical temperature $\beta_c^*(\alpha)$. Then, the ``critical inverse 
temperature'' $\beta_c(\alpha,\Lambda)$ of
the real system in a finite
volume $\Lambda$ can be infered to go to zero as $\beta_c(\alpha,\Lambda)\sim
\beta_c^*(\alpha)|\Lambda|^{-\delta}$.
It was numerically observed (\cite{CT1}, \cite{TA})
that $\beta_c^*(\alpha)$ depends weakly on $\alpha$,
which has lead the authors to conjecture that
all the systems with $0\leq \alpha\leq d$
have the same thermodynamic behaviour, i.e.
identical to the pure mean field case $\alpha=0$.
This ``universal'' mean field behaviour
was then given an intelligible explanation  
by Vollmayr-Lee and Luijten \cite{VL}.

\noindent 
We now wish to present an argument in favor of our conjecture, similar
in some ways to the strategy of \cite{VL}. Apart from helping 
to understand our
conjecture, it also sheds some light on the  weak dependence in $\alpha$
observed numerically in \cite{CT1}, \cite{TA}.
For simplicity we consider the case $q=2$, i.e. the Ising model.
Remember that Conjecture \ref{Conj1} claims that when 
$\phi$ is non-summable, then $\beta_{c}(\phi_N)\to 0$ as $N\to\infty$, where
$\beta_{c}(\phi_N)$ is the critical inverse 
temperature of the truncated model $\phi_N$.
Define
$$e_N:=\sum_{x\neq 0}\phi_N(x)=\sum_{x\in B_N}\phi(x)\,,$$
and consider the formal identity:
\begin{align}\nonumber
\beta\sum_{\substack{x\neq y}}
\phi_N(x-y)\sigma_x\sigma_y&=\beta e_N\sum_{\substack{x\neq y}}
\frac{\phi_N(x-y)}{e_N}\sigma_x\sigma_y\nonumber\\
&\equiv
\widehat{\beta}_N\sum_{\substack{x\neq y}}\widehat{\phi}_N(x-y)
\sigma_x\sigma_y\,,\nonumber
\end{align}
where we have defined the rescaled quantities
\begin{equation}\label{rescale}
\widehat{\beta}_N:=\beta e_N\,,\quad\quad
\widehat{\phi}_N(x):=\frac{\phi_N(x)}{e_N}\,.
\end{equation}
Since $\lim_Ne_N=+\infty$ (we assume $\phi$ is non-summable), we have 
$\lim_N\widehat{\beta}_N=+\infty$. 
Moreover, the new
potential $\widehat{\phi}_N$ has the following properties: 1) it has range
at most $N$, 2) it is summable
\begin{equation}\nonumber
\sum_{x\neq 0}\widehat{\phi}_N(x)=1\,,
\end{equation}
and 3)
$\lim_N\widehat{\phi}_N(x)=0$ for all $x$. Such
properties remind those of \textsf{Kac potentials}, 
which are of the form
$\Phi_\gamma(x):=\gamma^d\varphi(\gamma x)$, where $\varphi(x)\geq 0$ is 
bounded, supported by $[-1,+1]^d$, with $\int\varphi(x)\mathrm{d}x=1$,
and $\gamma>0$ is a small
\textsf{scaling parameter}. $\Phi_\gamma$ thus 
has the following properties: 1) it
has range $\gamma^{-1}$, 2) it is summable:
$$\int\Phi_\gamma(x)\mathrm{d}x=1\quad\forall\gamma>0\,,$$
and 3) $\lim_{\gamma\to 0^+}\Phi_\gamma(x)=0$ for all $x$. 
It is well known (\cite{KUH}, \cite{LP}) that 
such potentials give, in the \textsf{van der Waals limit} 
$\gamma\to 0^+$, a justification of
the van der Waals-Maxwell theory of liquid-vapor equilibrium: in this
limit, the properties of the system converge to those of mean
field, \emph{regardless of the details of the function $\varphi$}.
Moreover, it is known (\cite{CP}, \cite{BZ}) that for $\gamma>0$ the model
is a good approximation to mean field: a phase transition
occurs \emph{before} reaching the mean field regime, and
$\sup_{\gamma>0}\beta_c(\Phi_\gamma)<+\infty$.\\

\noindent It is tempting to ask whether this 
mean field behaviour of Kac potentials 
$\Phi_\gamma$ at small $\gamma$ also holds 
for our potential $\widehat{\phi}_N$ at large $N$, and to identify the 
parameters $\gamma^{-1}$ and $N$. 
Although $\widehat{\phi}_N$ is not obtained by a rescaling of a 
given function, as $\Phi_\gamma$ is, one 
can expect that for a
reasonable potential $\phi$ (for example of the type \eqref{potnonext}), 
the critical temperature 
$\beta_c(\widehat{\phi}_N)$ is uniformly bounded in $N$: 
$\sup_N\beta_c(\widehat{\phi}_N)<+\infty$. Therefore, 
if $N$ is large enough so that $\widehat{\beta}_N> 
\sup_N\beta_c(\widehat{\phi}_N)\geq
\beta_c(\widehat{\phi}_N)$, we have a phase transition in the truncated model
$\phi_N$, as predicted by our conjecture.
Moreover, since we expect that the properties of the system with 
$\widehat{\phi}_N$ converge to those of mean field when $N\to\infty$, 
\emph{independently} of the fine structure of $\phi$, this again 
is in favor of the non-dependence on $\alpha$ 
observed in \cite{CT1} and \cite{TA}.\\

\noindent Unfortunately,
the unique case where this scenario can be implemented rigorously is for the
constant potential: 
$\phi(x)=c>0$ for all $x\neq 0$ (or, equivalently, if $\alpha=0$). In this case
we have $\widehat{\phi}_N(\cdot)=|B_N|^{-1}1_{\|x\|\leq N}(\cdot)$,
and the correspondence $\gamma^{-1}\equiv N$ can be done (this case is treated 
in details in \cite{FrPf2}).\\
\paragraph{On the General Problem of Truncation.} We have seen that
in the long range ferromagnetic Potts model, 
non-summable interactions with the symmetry \eqref{symsimple}
are pathological in the sense that in the limit
$N\to\infty$,
\begin{equation}\nonumber
\mu_{\phi_N}^{\beta,s}\Rightarrow \delta_s\,
\end{equation}
for all $\beta>0$ and all $s\in\{1,2,\dots,q\}$.
A natural question is to ask whether this weak convergence also occurs in the 
case where $\phi$ \emph{is} sommable, i.e. when $\mu_{\phi}^{\beta,s}$
is well defined. Is it that
\begin{equation}\label{convfaible}
\mu_{\phi_N}^{\beta,s}\Rightarrow \mu_\phi^{\beta,s}\,
\end{equation}
in the limit $N\to\infty$? The answer 
to this question is negative in $d=1$. Namely, for the Ising model $q=2$, 
there exist summable potentials for which $\mu_\phi^{\beta,+}(\sigma_0=+1)>
\frac{1}{2}$ at low temperature (\cite{Dy},\cite{FrSp}), 
but $\mu_{\phi_N}^{\beta,+}(\sigma_0=+1)=0$ for 
all $N$ as well known (see Remark \ref{remunique}).
On can thus wonder if the weak convergence \eqref{convfaible}
occurs in dimensions $d\geq 2$. If this happens to be the case, 
it would lead to the following highly non-trivial
fact: assume $\phi$ is sommable and that a phase transition occurs in the 
untruncated system:
$$\mu_{\phi}^{\beta,s}(\sigma_0=s)>\frac{1}{q}\,.$$ Then, if \eqref{convfaible}
holds, there exists
an integer $N$ such that a phase transition occurs in the truncated model:
$$\mu_{\phi_N}^{\beta,s}(\sigma_0=s)>\frac{1}{q}\,.$$
That is, the tail of the interaction plays no important role in the occurence 
of phase transitions in summable long range models.
This question is again the analog of the general problem of truncation  
that has been posed in the context of long range percolation: if 
$P(0\leftrightarrow\infty)>0$, does there exist some large $N$ such that
$P_N(0\leftrightarrow\infty)>0$? This is believed to be true in 
general, with no assumption on the edge probability 
(other than, say, the symmetry \eqref{blob}, or for sparse interactions as
\eqref{blub}). This property is non-trivial for the following reason:
the truncated measure $P_N$ converges weakly to $P$, but the
Portmanteau Theorem (see \cite{Bill}) doesn't apply, due to the fact that in
the product topology, the boundary of the 
event $\{0\leftrightarrow\infty\}$ has strictly positive probability 
($1$, in fact). One can thus not conclude that
$P_N(0\leftrightarrow\infty)\to P(0\leftrightarrow\infty)$.
Our results of Section \ref{Smethod} and various 
existing results \cite{MS, Be, SSV, FLS}
give affirmative answer to the problem of truncation for particular
cases.\\

\noindent \textbf{Acknowledgments:} 
We
thank Vladas Sidoravicius for introducing us to the problem of truncation in 
long range independent percolation and for convincing us to publish the
results of Section \ref{Smethod}. 
We also thank Olle H\"aggstr\"om for suggesting to formulate these results 
for the Potts model.
S.Friedli 
thanks CBPF and IMPA for hospitality. B.N.B. de Lima thanks IMPA for 
partial financial support and various visits.

\bibliography{bibliochach2}

\begin{thebibliography}{10}

\bibitem{ACCN}
M.~Aizenman, J.~T. Chayes, L.~Chayes, and C.~N. Newman.
\newblock Discontinuity if the magnetization in one-dimensional $1/|x-y|^2$
  {Ising} and {P}otts models.
\newblock {\em Journ. Stat. Phys.}, 50:1--40, 1988.

\bibitem{AKN}
M.~Aizenman, H.~Kesten, and C.~N. Newman.
\newblock Uniqueness of the infinite cluster and continuity of the connectivity
  functions for short and long range percolation.
\newblock {\em Commun. Math. Phys.}, 111:505--531, 1987.

\bibitem{Be}
N.~Berger.
\newblock Transience, recurrence and critical behavior for long-range
  percolation.
\newblock {\em Commun. Math. Phys.}, 226:531--558, 2002.

\bibitem{Bill}
P.~Billingsley.
\newblock {\em Convergence of Probability Measures}.
\newblock Wiley, New York, 1968.

\bibitem{Bo}
T.~Bodineau.
\newblock Slab percolation for the {Ising} model.
\newblock {\em Probab. Theory and Rel. Fields}, 132(1):83--118, 2005.

\bibitem{BZ}
A.~Bovier and M.~Zahradn{\'\i}k.
\newblock The low-temperature phases of {Kac-{Ising}} models.
\newblock {\em Journ. Stat. Phys.}, 87:311--332, 1997.

\bibitem{CMT}
S.~A. Cannas, A.~de~Magalh{\~a}es, and F.~A. Tamarit.
\newblock Evidence of exactness of the mean field theory in the nonextensive
  regime of long-range spin models.
\newblock {\em Phys. Rev. Lett.}, 122:597--607, 1989.

\bibitem{CT1}
S.~A. Cannas and F.~A. Tamarit.
\newblock Long-range interactions and nonextensivity in ferromagnetic spin
  models.
\newblock {\em Phys. Rev. B}, 54:R12661--R12664, 1986.

\bibitem{CP}
M.~Cassandro and E.~Presutti.
\newblock Phase transitions in {Ising} systems with long but finite range
  interactions.
\newblock {\em Mark. Proc. Rel. Fields}, 2:241--262, 1996.

\bibitem{Dy}
F.~J. Dyson.
\newblock Existence of a phase-transition in a one-dimensional {Ising}
  ferromagnet.
\newblock {\em Commun. Math. Phys.}, 12:91--107, 1969.

\bibitem{FLS}
S.~Friedli, B.~N.~B. de~Lima, and V.~Sidoravicius.
\newblock On long range percolation with heavy tails.
\newblock {\em Elect. Comm. in Probab.}, 9:175--177, 2004.

\bibitem{FrPf2}
S.~Friedli and C.-{\'E}. Pfister.
\newblock Non-analyticity and the van der {W}aals limit.
\newblock {\em Journ. Stat. Phys.}, 114(3/4):665--734, 2004.

\bibitem{FrSp}
J.~Fr{\"o}hlich and T.~Spencer.
\newblock The phase transition in the one-dimensional {Ising} model with
  $1/r^2$ interaction energy.
\newblock {\em Commun. Math. Phys.}, 84:87--101, 1982.

\bibitem{Geo}
H.-O. Georgii.
\newblock {\em Gibbs Measures and Phase Transitions}.
\newblock de Gruyter Studies in Mathematics, 1988.

\bibitem{GHM}
H.-O. Georgii, O.~H{\"a}ggstr{\"o}m, and C.~Maes.
\newblock The random geometry of equilibrium phases.
\newblock In C.~Domb and J.~Lebowitz, editors, {\em Phase Transitions and
  Critical Phenomena Vol. 18}, pages 1--142. Academic Press, London, 2001.

\bibitem{GM}
G.~Grimmett and J.~M. Marstrand.
\newblock The supercritical phase of percolation is well behaved.
\newblock {\em Proc. Roy. Soc. London Ser}, A 430:439--457, 1990.

\bibitem{HNT}
P.~Hertel, H.~Narnhofer, and W.~Thirring.
\newblock Thermodynamic functions for fermions with gravostatic and
  electrostatic interactions.
\newblock {\em Commun. Math. Phys.}, 28:159--176, 1972.

\bibitem{KUH}
M.~Kac, G.~E. Uhlenbeck, and P.~C. Hemmer.
\newblock On the van der {W}aals theory of the vapor-liquid equilibrium.
\newblock {\em Journ. Math. Phys.}, 4:216--228, 1962.

\bibitem{Ke1}
H.~Kesten.
\newblock Asymptotics in high dimensions for percolation.
\newblock In O.~S. Publ., editor, {\em Disorder in physical systems}, pages
  219--240. Oxford Univ. Press, New York, 1990.

\bibitem{LP}
J.~L. Lebowitz and O.~Penrose.
\newblock Rigorous treatment of the van der {Waals-Maxwell} theory of the
  liquid-vapor transition.
\newblock {\em Journ. Math. Phys.}, 7:98--113, 1966.

\bibitem{LSS}
T.~M. Liggett, R.~H. Schonmann, and A.~M. Stacey.
\newblock Domination by product measures.
\newblock {\em Ann. Probab.}, 25:71--95, 1997.

\bibitem{MS}
R.~Meester and J.~E. Steif.
\newblock On the critical value for long range percolation in the exponential
  case.
\newblock {\em Commun. Math. Phys.}, 180:483--504, 1996.

\bibitem{PS1}
S.~A. Pirogov and Y.~G. Sinai.
\newblock Phase diagrams of classical lattice systems.
\newblock {\em Teoreticheskaya i Matematicheskaya Fizika}, 26(1):61--76, 1976.

\bibitem{SSV}
V.~Sidoravicius, D.~Surgailis, and M.~E. Vares.
\newblock On the truncated anisotropic long-range percolation on
  $\mathbf{Z}^2$.
\newblock {\em Stoch. Proc. and Appl.}, 81:337--349, 1999.

\bibitem{TA}
F.~Tamarit and C.~Anteneodo.
\newblock Rotators with long-range interactions: Connections with the
  mean-field approximation.
\newblock {\em Phys. Rev. Lett.}, 84:208, 2000.

\bibitem{VL}
B.~P. Vollmayr-Lee and E.~Luijten.
\newblock {Kac}-potential treatment of nonintegrable interactions.
\newblock {\em Phys.Rev. E}, 63:031108, 2001.

\end{thebibliography}


\begin{thebibliography}{xxxxx}

\bibitem[AN]{AN} Aizenman M., Newman C.M., \emph{Discontinuity of the
Percolation Density in One-Dimensional $1/|x-y|^s$ Percolation
Models}, Commun. Math. Phys. {\bf 107}, 611-647 (1986).
\bibitem[Bi]{Bi} Billingsley P., \emph{Convergence of Probability Measures},
Jonh Wiley , New York, 1968.
\bibitem[HS]{HS} Hara T., Slade G., \emph{Mean-Field Critical Behavior
 for Percolation
in High Dimensions}, Commun. Math. Phys. {\bf 128}, 333-391
(1990).
\bibitem[NS]{NS} Newman C.M., Schulman L.S., \emph{One Dimensional $1/|i-j|^s$
Percolation Models: The Existence of a Transition for $s\leq 2$}, Commun. Math.
Phys. {\bf 104}, 547-571 (1986).
\bibitem[SSV]{SSV} Sidoravicius V., Surgailis D., Vares M.E., \emph{On the
Truncated Anisotropic Long-Range Percolation on $\mathbf{Z}^2$}, Stoch. Proc. 
and Appl. {\bf 81}, 337-349 (1999).
\end{thebibliography}

\end{document}